\documentclass[12pt]{iopart}
\usepackage{iopams}
\usepackage[square,numbers,sort&compress]{natbib}
\usepackage{graphicx}
\usepackage{dcolumn}
\usepackage{bm}
\usepackage[dvipsnames]{xcolor}

\begin{document}

\sloppy

\title[Quantifying the impact of vibrational nonequilibrium in plasma catalysis]{Quantifying the impact of vibrational nonequilibrium in plasma catalysis: Insights from a molecular dynamics model of dissociative chemisorption}

\author{Kristof M Bal and Erik C Neyts}
\address{Department of Chemistry and NANOlab Center of Excellence, University of Antwerp, Universiteitsplein 1, 2610 Antwerp, Belgium}
\ead{kristof.bal@uantwerpen.be}

\date{\today}

\begin{abstract}
The rate, selectivity and efficiency of plasma-based conversion processes is strongly affected by nonequilibrium phenomena.
High concentrations of vibrationally excited molecules are such a plasma-induced effect.
It is frequently assumed that vibrationally excited molecules are important in plasma catalysis because their presence lowers the apparent activation energy of dissociative chemisorption reactions and thus increases the conversion rate.
A detailed atomic-level understanding of vibrationally stimulated catalytic reactions in the context of plasma catalysis is however lacking.
Here, we couple a recently developed statistical model of a plasma-induced vibrational nonequilibrium to molecular dynamics simulations, enhanced sampling methods, and machine learning techniques.
We quantify the impact of a vibrational nonequilibrium on the dissociative chemisorption barrier of H$_2$ and CH$_4$ on nickel catalysts over a wide range of vibrational temperatures.
We investigate the effect of surface structure and compare the role of different vibrational modes of methane in the dissociation process.
For low vibrational temperatures, very high vibrational efficacies are found, and energy in bend vibrations appears to dominate the dissociation of methane.
The relative impact of vibrational nonequilibrium is much higher on terrace sites than on surface steps.
We then show how our simulations can help to interpret recent experimental results, and suggest new paths to a better understanding of plasma catalysis.
\end{abstract}

\maketitle

\section{Introduction}

Plasma-based processes are considered to be a promising approach to gas conversion, such as CO$_2$ dissociation, methane reforming, and nitrogen fixation~\cite{Adamovich2017,Bogaerts2018}.
The huge potential of plasma technology can be traced to two key arguments.
On one hand, plasma is generated by electric power and a plasma process can be quickly turned on or off, making it suitable in the context of an intermittent electricity supply from renewable sources.
On the other hand, nonthermal technological plasmas can combine high reactivity with near-ambient operating temperatures and pressures.
As a result, plasma conversion processes have high conversion rates and, most intriguingly, can shift the equilibrium of thermodynamically unfavorable reactions towards products of interest~\cite{vanRooij2015,Uytdenhouwen2021}.

Plasma catalysis, which combines nonthermal plasma technology with a heterogeneous catalyst, is in particular receiving increasing attention~\cite{Neyts2015,Liu2020,Bogaerts2020,Zhang2021}.
Indeed, it is a quite tantalizing prospect to combine the highly reactive plasma chemistry with the kinetic activity and selectivity of a catalyst.
Yet, a fundamental understanding of plasma catalysis is still lacking.
Most notably, the microscopic plasma-induced phenomena at the catalyst surface remain elusive, as well as their potential impact on the activity of the catalyst~\cite{Whitehead2016}.
Only very recently, experimental methods have been devised that can obtain in situ insights on the surface processes and intermediates in plasma-catalytic setups~\cite{Zhang2020,Winter2020}.

Several initial insights have come from modeling.
The reactivity of plasma-generated radicals on the catalyst surface can be rather straightforwardly simulated in molecular dynamics (MD) simulations~\cite{Somers2012} and also incorporated in microkinetic models~\cite{Engelmann2020}.
Moreover, quantum chemical calculations have predicted that the strong electric fields and surface charges found in packed bed plasma reactors~\cite{Zhang2018,Kruszelnicki2020} can shift the acid--base properties of surfaces, and selectively tune adsorption strengths~\cite{Bal2018,Jafarzadeh2020}.

The role of vibrational nonequilibrium is somewhat controversial in the context of plasma catalysis.
In many plasmas the average energy stored in specific vibrational modes is higher than the average thermal energy: The average vibrational temperature of specific chemically active normal modes can be well above 1000~K in plasmas that are otherwise at room temperature.
In addition, beam studies have suggested that dissociative chemisorption of methane is promoted more efficiently by vibrational energy than translational energy~\cite{Beck2003,Smith2004,Juurlink2005}.
As a result, reactions of vibrationally excited molecules have been postulated to be important in a wide range of plasma-catalytic processes, including CO$_2$ splitting, CH$4$ reforming, and N$_2$ activation~\cite{Nozaki2013,Kim2016,Mehta2018,Sheng2018,Rouwenhorst2019,Sheng2020,Engelmann2020}.
However, no detailed mechanistic evidence is available that conclusively supports these claims.

Recently, we have introduced a model for vibrationally stimulated reactions, based on MD simulations and free energy methods~\cite{Bal2020JPCL}. 
Our method is specifically designed to transparently mimic a vibrational nonequilibrium as found in plasmas.
This is achieved by modeling the vibrational nonequilibrium as a perturbed canonical ensemble, characterized by two different temperatures and energy distributions.
Rate enhancement is then quantified in terms of changes in the apparent free energy barrier of the process of interest.
For a number of simple gas phase reactions, we were able to correctly predict rate enhancements and mode selectivities~\cite{Bal2020JPCL}.

In this contribution, we apply our ensemble-based method to quantify the impact of vibrational nonequilibrium on CH$_4$ dissociation on nickel catalysts.
In particular, we compare the vibrational efficacy of different classes of vibrational modes (symmetric stretch, asymmetric stretch, bend) on the reaction over different Ni surface facets.
In order to properly reconstruct free energy barriers within the limited MD time scale, we use advanced enhanced sampling methods~\cite{Laio2002,Barducci2008,Valsson2014} combined with machine learning techniques~\cite{Mendels2018}.

In section~\ref{sec:methods}, we review our methodology.
We then apply it first to the simpler H$_2$ dissociation reaction (section~\ref{sec:h2}), before moving to CH$_4$ dissociation (section~\ref{sec:ch4}).

\section{Methodology}
\label{sec:methods}

\subsection{Ensemble perspective on vibrationally stimulated reactions}

We have recently developed a simulation approach that treats chemical reaction under vibrational nonequilibrium in an averaged manner~\cite{Bal2020JPCL}.
There are two central assumptions to this methodology:
\begin{enumerate}
  \item The effect of vibrational excitation on the reactivity of a molecule can be fully attributed to configurational effects.
  That is, because of the excitation, the probability of a molecule to be in a ``reactive'' configuration is altered.
  This change in likelihood can in turn be expressed as an apparent change (usually reduction) of the reaction barrier.
  \item Vibrational excitation is an ensemble property.
  Each degree of freedom $i$ in the considered system is equilibrated at a temperature $T_i$.
  For vibrational modes this means that quantization is not considered, or, put differently, we simply assume that all the populations of all levels are Boltzmann-distributed, according to $T_i$.
  Any property calculated for the system is thus valid for the particular ensemble of molecular states, defined by $\{ T_i \}$.
\end{enumerate}
In practice, we simulate systems for which most degrees of freedom have a low background temperature $T_\mathrm{bg}$, and only a limited number of vibrational modes are at a higher vibrational temperature $T_\mathrm{vib}$.

In an MD simulation, it is customary to apply a thermostat to enforce a constant temperature $T$ to all modes.
We can still do this in our approach, where we use the thermostat to give all modes the temperature $T = T_\mathrm{bg}$.
A selected degree of freedom $u$ can then be ``excited'' by applying an external bias potential $V (u)$ that modifies the probability distribution $p (u)$, as if $u$ is at a higher temperature $T_\mathrm{vib}$---and this while a thermostat is still only giving a kinetic energy corresponding to $T = T_\mathrm{bg}$ to all modes. 
For example, if $u$ is a bond length, we would expect $p(u)$ to become broader and shifted to larger values at higher temperature.
This change in $p(u)$ may then affect the barrier height of a chemical reaction of interest, which can be quantified by calculating the reaction free energy profile with and without the application of $V (u)$, at a simulated temperature of $T = T_\mathrm{bg}$.

In practice, the full workflow is as follows:
\begin{enumerate}
  \item Identify a reaction coordinate $\chi (\mathbf{R})$ for the reaction of interest, and a parametrization of the considered vibrational mode $u (\mathbf{R})$, both expressed as a function of the system coordinates $\mathbf{R}$.
  \item Sample the distribution $p(u)$ at a high temperature $T = T_\mathrm{vib}$. 
  This can be done by simulating a single molecule at a temperature $T_\mathrm{vib}$, i.e., considering a system for which the thermostat temperature $T = T_\mathrm{vib}$.
  It is this approach that introduces the assumption of a Boltzmann-distributed vibrational nonequilibrium: A standard, canonical thermostat will naturally sample from a Boltzmann distribution and, as a result, produce a $p (u)$ that obeys it.
  \item Parametrize a bias potential $V (u)$ that enforces the distribution $p (u)$ at a temperature $T = T_\mathrm{bg}$.
  This is done by using the variationally enhanced sampling~\cite{Valsson2014} algorithm, using $p (u)$ as its target distribution.
  The thermostat temperature is now $T = T_\mathrm{bg}$.
  \item Simulate the full system at $T = T_\mathrm{bg}$, while applying $V (u)$ as an external bias, and reconstruct the free energy surface $F (\chi)$ using an enhanced sampling method of choice, here well-tempered metadynamics~\cite{Laio2002,Barducci2008}.
\end{enumerate}

The advantage of this approximation is that we can describe the impact of vibrational excitation in manner fully analogous to how we would treat reactions under conventional thermal equilibrium conditions.
That is, the reactivity of the gas is described in terms of the probability density along a reaction coordinate.
The probability correlates to the free energy barrier and rate of the reaction, and is an average property of an ensemble of gas molecules at a specified temperature.
By applying the bias $V(u)$ we simulate a different ensemble, corresponding to a higher $T_\mathrm{vib}$ in $u$, while characterizing the reactivity of the gas in a completely analogous manner, calculating the same thermodynamic and kinetic observables.

The key disadvantage of the approximation is that we do not explicitly the quantized nature of the vibrational energy and, as a result, do not independently treat the reactivity of specific excited states.
A more fine-grained treatment of the kinetics of vibrationally excited states, which also allows for a self-consistent derivation of the vibrational distribution function (rather than assuming a Boltzmann distribution), is possible with global kinetic models.
The literature on such models is rather bountiful for many types of CO$_2$ plasma~\cite{Kozak2014,Capitelli2017,Silva2018,Viegas2019}, but much less so for CH$_4$~\cite{Sun2020,Heijkers2020}.

\subsection{Reaction coordinates and enhanced sampling}

Gas--surface reactions are rather complex due to the many degrees of freedom involved.
As a result, finding the reaction coordinate $\chi = \chi (\mathbf{R})$ is a nontrivial task.
Herein, we express $\chi$ as a linear combination of several simpler building blocks $\{ s_i (\mathbf{R}) \}$.
The linear coefficients of this expansion are found using harmonic linear discriminant analysis (HLDA), a machine learning method for the identification of reaction coordinates~\cite{Mendels2018}.
In HLDA, one assumes that the reaction of interest proceeds between two known metastable states $A$ and $B$.
Short MD simulations are then run in both states, and the linear coefficients in $\chi$ are calculated from the fluctuations of all $s_i$ in each state.

All considered $s_i$ are physically intuitive degrees of freedom.
One of them is the longest bond in the reacting molecule which, if stretched enough, also corresponds to the bond that breaks upon dissociative chemisorption.
It is calculated from all molecular contacts $r_j$ as
\begin{equation}
  s_\mathrm{bond} = \gamma \ln \sum_j e^{r_j / \gamma} ,
\end{equation}
in which $\gamma = 0.01$~\AA{}.
For methane, we consider all C--H contacts, whereas for H$_2$, this expression reduces to the H--H bond length.

Another degree of freedom is the average coordination number of C or H with the metal surface atoms, which describes the binding of the molecule on the surface.
For all (C,H)--Ni distances $r_j$, it is defined as
\begin{equation}
  s_\mathrm{(C,H)-Ni} = \frac{1}{N} \sum_j^N \frac{1 - (r_j/r_0)^6}{1 - (r_j/r_0)^{12}} ,
\end{equation}
in which $r_0 = 2.5$~\AA{} for $s_\mathrm{C-Ni}$ and $r_0 = 2.0$~\AA{} for $s_\mathrm{H-Ni}$, respectively.

Finally, dissociation of CH$_4$ also requires a distortion of the tetrahedral shape of the molecule $s_\mathrm{tet}$, which we measure from all possible bond angles $\theta_j$ by counting those close to the ideal tetrahedral angle of $\theta \approx 109.5^\circ$.
To calculate this quantity, we use kernel density estimation as implemented in the PLUMED package~\cite{Tribello2014,PLUMED2019}.

As noted, we use two different enhanced sampling approaches.
In order to generate an external bias $V (u)$ to enforce nonequilibrium, we use variationally enhanced sampling (VES)~\cite{Valsson2014}.
VES iteratively generates a bias potential $V (u)$ so that the probability distribution $p(u)$ matches some pre-defined distribution chosen by the user.
In its standard form, as used here, VES expands $V (u)$ as linear combination of basis functions (we choose Chebyshev polynomials) so that bias optimization entails the refinement of the respective linear expansion coefficients.

While $p(u)$ is an input function for VES, $p (\chi)$ is unknown and the quantity we ultimately wish to obtain.
This is because $p (\chi)$ is directly related to the reaction free energy surface $F (\chi)$, through the relation
\begin{equation}
  F (\chi) = - k_B T \ln p (\chi) ,
\end{equation}
in which $k_B$ is the Boltzmann constant.
If high free energy barriers are present along $F (\chi)$, $p (\chi)$ cannot be properly sampled within the time scale of MD simulations.
Therefore, we must enhance the exploration of configuration space along $\chi$.
We do this by using metadynamics~\cite{Laio2002,Barducci2008}.
In metadynamics, an external bias potential $V (\chi)$ is slowly generated by periodically adding small repulsive Gaussian kernels at time steps $t_i$, centered around $\chi (t_i)$.
This way, deep free energy minima are gradually filled, and exploration of configuration space is enhanced in a targeted manner.
In order to reconstruct the original $p (\chi)$ from the biased metadynamics simulation, one can use reweighting techniques to remove \textit{a posteriori} the effect of the bias potential so as to recover $F (\chi)$ and free energy barriers~\cite{Tiwary2015,Bal2020JCP}.
The same sampling strategy is used for each $T_\mathrm{vib}$, so that gas reactivity at different degrees of vibrational nonequilibrium is quantified as an ensemble property in a consistent framework.

\subsection{Simulation details}

All simulations were carried out using a combination of two simulation codes: LAMMPS~\cite{Plimpton1995} to evaluate interatomic forces and integrate the equations of motion in the MD simulations, and the PLUMED plugin~\cite{Tribello2014,PLUMED2019} to apply bias potentials, analyze probability distributions, and calculate free energies.
Interatomic interactions were described using a ReaxFF force field~\cite{vanDuin2001} parameterized for Ni/C/H systems~\cite{Mueller2010}.
Equations of motion were integrated with a time step of 0.1~fs and temperature control ($T_\mathrm{bg}$) was achieved using a generalized Langevin equation (GLE) thermostat optimized for efficient sampling~\cite{Ceriotti2009,Ceriotti2010}.
For each studied vibrational mode $i$, a suitable representation $u_i (\mathbf{R})$ was designed (see below); $p(u)$ was sampled for each $T = T_\mathrm{vib}$ over 100~ps, and $V(u)$ was parametrized over 1~ns.
For each combination of reaction and surface facet, a reaction coordinate was $\chi$ parametrized at $T = T_\mathrm{bg} = 500$~K from two 25~ps MD runs in the undissociated and dissociated state, respectively.
A free energy surface was derived using well-tempered metadynamics~\cite{Laio2002,Barducci2008} and reweighting techniques~\cite{Tiwary2015,Bal2020JCP} over 1~ns, at the same temperature.
For each condition, the free energy surface was calculated in three independent runs so that error bars can be reported on the barriers.

Sample inputs to reproduce the reported simulations are deposited on PLUMED-NEST, the public repository of the PLUMED consortium~\cite{PLUMED2019}, as plumID:21.013~\cite{data}.

\section{Results and discussion}

\subsection{Hydrogen dissociation}
\label{sec:h2}

As a first step, we consider a dissociative chemisorption process in which only one internal degree of freedom can be excited.
The reaction H$_2 \rightarrow 2$~H* involves a subset of the elements in the more complex dissociation reaction of CH$_4$; the only vibrational mode in H$_2$ is the H--H stretch.
As such, the reaction also shares some characteristics with the industrially important N$_2$ activation process.

In order to have a well-defined domain $u \in [0,1]$, we define
\begin{equation}
  u_\mathrm{H-H} = \frac{1 - \left ( \frac{r_\mathrm{H-H} - d_0}{r_0} \right )}{1 - \left ( \frac{r_\mathrm{H-H} - d_0}{r_0} \right )^2} ,
\end{equation}
in which $d_0 = 0.25$~\AA{} and $r_0 = 0.5$~\AA{}.
We investigate a range of vibrational temperatures $T_\mathrm{vib}$ between 750~K and 2500~K, in intervals of 250~K.

\begin{table}
\centering
\caption{\label{tab:h2} H$_2$ dissociation on different nickel surface facets.
Expansion coefficients of the reaction coordinate $\chi = c_1 s_\mathrm{H-H} + c_2 s_\mathrm{H-Ni}$, activation free energy barrier $\Delta^\ddagger F$ (in kcal/mol) at thermal equilibrium ($T = T_\mathrm{bg}$), average vibrational efficacy $\overline{\alpha}$ at low $T_\mathrm{vib}$ and extrapolated vibrational efficacy $\alpha_\infty$.}
\begin{tabular}{lcccc}
\br
facet          & $c_1,c_2$  & $\Delta^\ddagger F$  & $\overline{\alpha}$ & $\alpha_\infty$\\
\mr
(111)          & 0.992,0.124                       & $14.6 \pm 0.3$       & $2.16 \pm 0.18$     & $1.05 \pm 0.01$ \\
(001)          & 0.990,0.141                       & $13.9 \pm 0.1$       & $2.13 \pm 0.16$     & $1.13 \pm 0.18$ \\
(211)          & 0.993,0.118                       & $13.1 \pm 0.3$       & $2.12 \pm 0.17$     & $1.01 \pm 0.07$ \\
\br
\end{tabular}
\end{table}

The reaction coordinate $\chi$ was defined as a linear combination of $s_\mathrm{H-H}$ and $s_\mathrm{H-Ni}$ only; to illustrate, the expansion coefficients are given in table~\ref{tab:h2} for each investigated reaction.
On all facets, the stretching of the H--H bond dominates the reaction coordinate, while the coordination with the metal atoms still plays a sizeable role---note that $s_\mathrm{H-Ni}$ takes values between 0 and 6, while $s_\mathrm{H-H}$ is restrained to values up to 1.5~\AA{}.
We first compute the free energy surface at thermal equilibrium, i.e., with all modes at $T = T_\mathrm{bg}$.
At 500~K, we find that free energy barriers $\Delta^\ddagger F$ are quite similar (within $k_B T$), but suggest a reactivity trend (111)~$<$ (001)~$<$ (211).

\begin{figure}[t]
\centering
\includegraphics{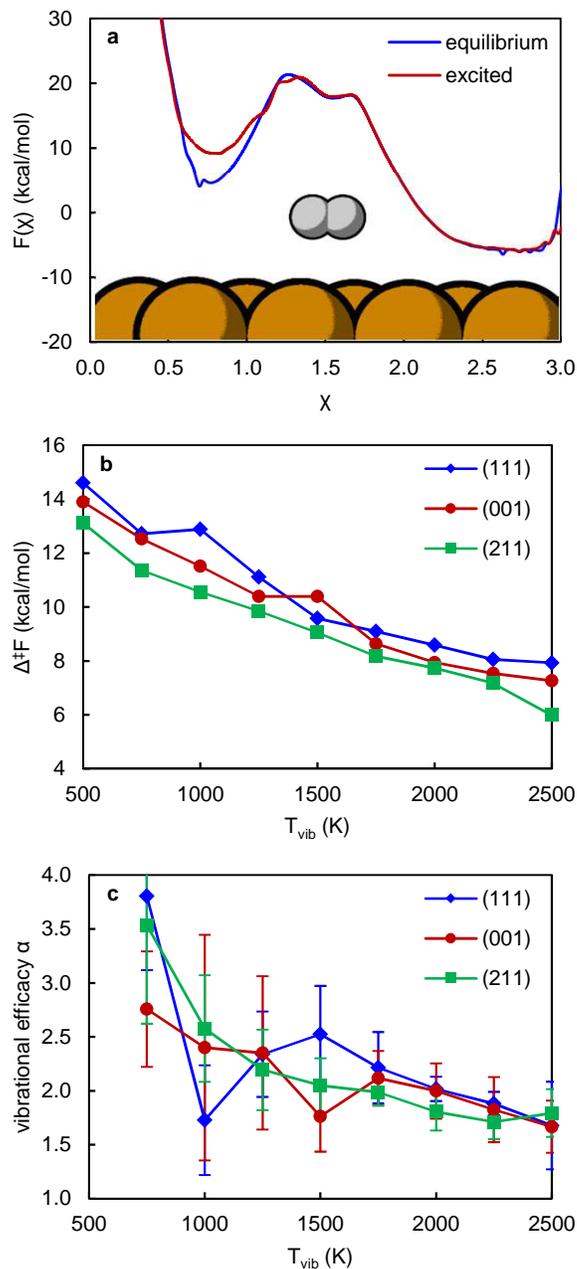}
\caption{\label{fig:panel-h2} H$_2$ dissociation on nickel surfaces at $T_\mathrm{bg} = 500$~K.
(a) Effect of vibrational excitation on the free energy surface $F (\chi)$ of dissociative adsorption on the Ni(111) surface (inset): comparing thermal equilibrium ($T_\mathrm{vib} = 500$~K) with $T_\mathrm{vib} = 1500$~K.
(b) Apparent free energy barrier $\Delta^\ddagger F^*$ and (c) vibrational efficacy $\alpha$ as a function of $T_\mathrm{vib}$ for the different facets.
Lines are drawn to guide the eye.}
\end{figure}

The effect of increasing $T_\mathrm{vib}$ on the free energy surface $F(\chi)$ is depicted in figure~\ref{fig:panel-h2}a for the reaction on the (111) facet.
At $T_\mathrm{vib} = 1500$~K (and $T_\mathrm{bg} = 500$~K) the free energy of H$_2$ vapor is shifted up relative to thermal equilibrium at 500~K; the dissociated state remains unaffected.
As a result, the apparent dissociation barrier is decreased.

In figure~\ref{fig:panel-h2}b, we plot the apparent free energy barrier $\Delta^\ddagger F$ as a function of $T_\mathrm{vib}$ on the three considered facets.
In all cases, $\Delta^\ddagger F$ decreases monotonically with increasing $T_\mathrm{vib}$.
Noting that the reaction rate has an exponential dependence on the barrier, we can see that even a modest vibrational nonequilibrium has a substantial impact.
Already at $T_\mathrm{vib} = 1000$~K the rate is predicted to increase by an order of magnitude.

We can now proceed to study the efficacy of this rate modulation in more detail.
A common formula used to incorporate vibrational nonequilibrium in chemical kinetics models is the Fridman--Macheret (FM) $\alpha$ model~\cite{Fridman2008}.
Here, the apparent lowered barrier $\Delta^\ddagger F^*$ is expressed as a function of the original barrier $\Delta^\ddagger F$ and the excess vibrational energy in a mode of interest $E_\mathrm{vib}$:
\begin{equation}
  \Delta^\ddagger F^* = \Delta^\ddagger F - \alpha E_\mathrm{vib} .
\end{equation}
Here, $\alpha$ is the vibrational efficacy: the fraction of the excess vibrational energy that can be used to overcome the reaction barrier $\Delta^\ddagger F$ by apparently lowering it.
In simple gas phase reactions with reaction energy $\Delta F$ the FM model stipulates that the parameter $\alpha$ can be estimated from
\begin{equation}
  \alpha =\frac{\Delta^\ddagger F}{2\Delta^\ddagger F - \Delta F} , \label{eq:FM}
\end{equation}
implying that exergonic reactions are less stimulated than endergonic ones, that $\alpha \geq 0$, and, because $\Delta^\ddagger F \geq \Delta F$, that $\alpha \leq 1$.
Recent microkinetic modeling studies of plasma-catalytic processes have postulated the validity of Eq.~\eref{eq:FM} also for gas--surface reactions~\cite{Mehta2018,Engelmann2020}.

In principle, we can determine $\alpha$ directly from our estimated $\Delta^\ddagger F^*$:
\begin{equation}
  \alpha =\frac{\Delta^\ddagger F - \Delta^\ddagger F^*}{k_B (T_\mathrm{vib} - T_\mathrm{bg})} , \label{eq:eff}
\end{equation}
and plot the estimated $\alpha$ values in figure~\ref{fig:panel-h2}c.
We observe that our computed $\alpha$ is consistently larger than unity.
At $T_\mathrm{vib} = 750$~K, we even see $\alpha > 3$ on the (111) and (211) facet.
(Note, however, that $\alpha$ is very sensitive to variations of the barrier values at low $T_\mathrm{vib}$.)
$\alpha$ decreases with increasing vibrational temperature, but even at $T_\mathrm{vib} = 2500$~K, $\alpha > 1.5$ on all facets.

In order to compare the three facets, we report numerical values of $\alpha$ for each in table~\ref{tab:h2}.
We use two measures.
One is $\overline{\alpha}$, averaged for $T_\mathrm{vib} \in [1000,2000]$~K.
The other, $\alpha_\infty$, is obtained from the extrapolation $T_\mathrm{vib} \rightarrow \infty$.
It can be seen that for either definition of $\alpha$ there is no significant difference between the facets.
If we briefly return to a quantized picture of vibrational motion, a very high vibrational temperature implies that all molecules are excited to higher vibrational levels.
As a result, $\alpha_\infty \approx 1$ suggests that the H--H stretching mode fully couples to the reaction coordinate and all vibrational energy can be used to overcome the dissociation barrier.
This is in contrast to the FM picture, which would predict $\alpha < 0.5$ because the reaction is exergonic.

Vibrational efficacy is much higher at lower $T_\mathrm{vib}$.
Our model offers a classical explanation for this observation.
Suppose that, in order to carry out the reaction of interest and cross the transition state $\chi = \chi^*$, the vibrational degree of freedom must be in a reactive configuration $u = u^*$.
Let us assume that any rate enhancement is fully proportional to the probability to visit this state, i.e., that $u$ fully couples to $\chi$.
In the isolated molecule the state $u = u^*$ will have a potential energy that is $\Delta U$ higher than the potential energy of the equilibrium state.
By increasing $T$ from $T_\mathrm{bg}$ to $T_\mathrm{vib}$, the probability $p(u^*)$ to visit this state (relative to the equilibrium configuration) will then increase by a factor 
\begin{equation}
  \textrm{rate enhancement (our model)} \propto \exp \left( \frac{\Delta U}{k_B} \left( \frac{1}{T_\mathrm{bg}} - \frac{1}{T_\mathrm{vib}} \right) \right) .
\end{equation}
In contrast, when assuming a simple FM picture of rate enhancement (and also assuming full coupling to the reaction coordinate, so $\alpha = 1$) one has
\begin{equation}
  \textrm{rate enhancement (FM model)} = \exp \left( \frac{T_\mathrm{vib}-T_\mathrm{bg}}{T_\mathrm{bg}} \right) .
\end{equation}
As a result, if $\Delta U$ is large (corresponding to a stiff bond), the rate enhancement will increase very strongly at fairly low $T_\mathrm{vib}$, more strongly than accounted for in the FM model.
Therefore, $\alpha \gg 1$ if we define $\alpha$ as in the FM model.
In our model, the rate enhancement naturally plateaus at high $T_\mathrm{vib}$, which is reasonable: A process cannot be infinitely accelerated simply by virtue of having $E_\mathrm{vib} \gg \Delta^\ddagger F$.
The FM model, however, does not explicitly account for such a situation.

\subsection{Methane dissociation}
\label{sec:ch4}

The dissociative chemisorption of methane, CH$_4$~$\rightarrow$ CH$_3$*~+ H*, is a significantly more complex process than the dissociation of H$_2$.
Most importantly, the molecule possesses many more vibrational degrees of freedom (9 in total), several of which are degenerate.

We opt to lump all vibrational modes in three classes, each with their own choice of $u$.
Stretch modes are described by calculating the following descriptor for each C--H bond:
\begin{equation}
  \sigma_i = \frac{1 - \left ( \frac{r_i - d_0}{r_0} \right )}{1 - \left ( \frac{r_i - d_0}{r_0} \right )^2} ,
\end{equation}
in which we choose $d_0 = 0.85$~\AA{} and $r_0 = 0.25$~\AA{}.
The degree of freedom that best describes symmetric stretches is the average of all $\sigma_i$, whereas the three asymmetric stretches are lumped together as the standard deviation of all the $\sigma_i$.
We lump all five bending degrees of freedom together, and choose the $s_\mathrm{tet}$ descriptor also as the respective $u$.

\begin{table}
\centering
\caption{\label{tab:ch4} CH$_4$ dissociation on different nickel surface facets.
Activation free energy barrier $\Delta^\ddagger F$ (in kcal/mol) at thermal equilibrium ($T = T_\mathrm{bg}$) and average vibrational efficacy $\overline{\alpha}$ at low $T_\mathrm{vib}$ for the symmetric (s), asymmetric (as) and bend (b) modes, respectively.}
\begin{tabular}{lcccc}
\br
facet          & $\Delta^\ddagger F$  & $\overline{\alpha}_\mathrm{s}$ & $\overline{\alpha}_\mathrm{as}$ & $\overline{\alpha}_\mathrm{b}$ \\
\mr
(111)          &  $13.6 \pm 0.1$       & $2.40 \pm 0.20$               & $0.74 \pm 0.10$                 & $0.78 \pm 0.12$ \\
(001)          &  $13.5 \pm 0.4$       & $2.68 \pm 0.23$               & $0.88 \pm 0.02$                 & $0.82 \pm 0.11$ \\
(211)          &  $10.8 \pm 0.1$       & $1.70 \pm 0.14$               & $0.66 \pm 0.07$                 & $0.53 \pm 0.10$ \\
\br
\end{tabular}
\end{table}

We again optimize a reaction coordinate $\chi$ for each surface facet as an expansion of $s_\mathrm{C-H}$, $s_\mathrm{C-Ni}$, $s_\mathrm{H-Ni}$, and $s_\mathrm{tet}$.
For the reaction at each facet, we report in table~\ref{tab:ch4} the free energy barrier $\Delta^\ddagger F$ at thermal equilibrium ($T_\mathrm{vib} = T_\mathrm{bg} = 500$~K).
Recently, Engelmann et al. also assumed $T_\mathrm{bg} = 500$~K and Boltzmann-distributed vibrational levels (at $T_\mathrm{vib} = 1500$~K) in a microkinetic modeling study of plasma-catalytic nonoxidative coupling of methane~\cite{Engelmann2020}.
Here, we notice that our ReaxFF-based methodology cannot replicate the difference in reactivity of the Ni(111) and Ni(001) facets.
According to a recent analysis of experimental reaction barriers, dissociation on the latter facet should have a barrier that is about 6~kcal/mol lower than on the former~\cite{Sharada2017}.
The force field does, however, correctly predict that dissociation is more facile on the stepped (211) facet.

\begin{figure}[t]
\centering
\includegraphics{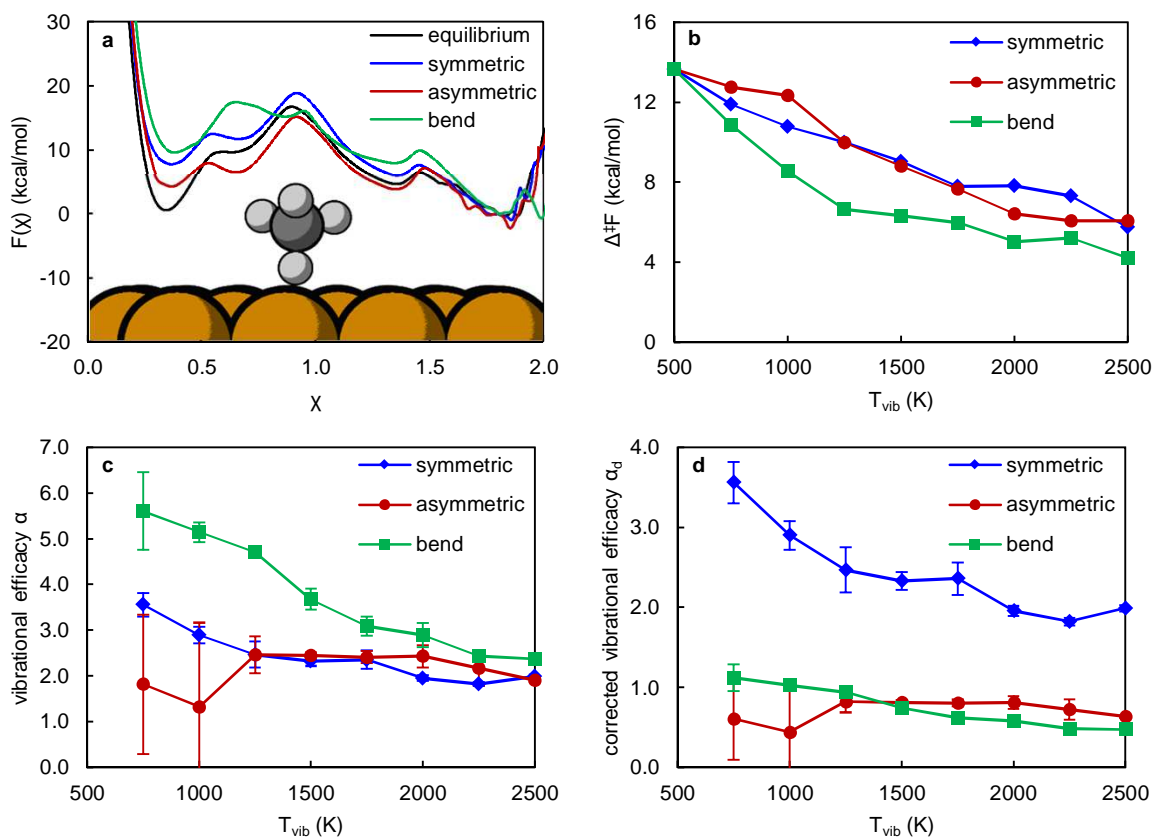}
\caption{\label{fig:panel-ch4} CH$_4$ dissociation on the Ni(111) surface at $T_\mathrm{bg} = 500$~K.
(a) Effect of vibrational excitation of different classes of CH$_4$ vibrational on the free energy surface $F (\chi)$ of dissociative adsorption (inset): comparing thermal equilibrium ($T_\mathrm{vib} = 500$~K) with $T_\mathrm{vib} = 1500$~K.
(b) Apparent free energy barrier $\Delta^\ddagger F^*$.
(c) Vibrational efficacy $\alpha$ as a function of $T_\mathrm{vib}$ for the different modes.
(d) Vibrational efficacy $\alpha_d$ taking into account degeneracy.
Lines are drawn to guide the eye.}
\end{figure}

In figure~\ref{fig:panel-ch4}a we plot the free energy surface $F (\chi)$ of CH$_4$ dissociative chemisorption on Ni(111).
We compare the thermal equilibrium case with vibrational nonequilibria at $T_\mathrm{vib} = 1500$~K involving the symmetric, asymmetric, and bending modes, respectively.
It can be seen that impact of excitation on the apparent barrier differs depending on the class of excited modes.
As was the case for H$_2$ dissociation, excitation of the stretch modes shifts the free energy of the gaseous state to higher values thus lowering the apparent barrier.
Excitation of bending modes, however, also changes the shape of $F (\chi)$ and shifts the the location of the transition state towards the reactant state.

In figure~\ref{fig:panel-ch4}b, we plot the apparent free energy barrier $\Delta^\ddagger F$ on the (111) facet, as a function of $T_\mathrm{vib}$ for the three considered classes of modes.
Here, too, we observe a difference in impact between excited stretch modes on one hand, and the bending modes on the other.
At low $T_\mathrm{vib}$, excitation of the bending modes will lead to a much larger decrease of the dissociation barrier (and increase of the rate) than the same excitation of either type of stretch mode.
This difference can be up to 4~kcal/mol, or a two orders of magnitude rate enhancement.
The efficacy of the bending modes, as calculated from Eq.~(\ref{eq:eff}), is therefore the highest, as can be seen in figure~\ref{fig:panel-ch4}c.
This does not, however, take into account degeneracy.
Indeed, at a given $T_\mathrm{vib}$, five times more excess energy is stored in the bending modes than in the symmetric stretch due to former's degeneracy.
Similarly, there are three asymmetric stretch modes.
For a more fair comparison, we can therefore correct $\alpha$ for the degeneracy $d$:
\begin{equation}
  \alpha_d = \frac{1}{d} \frac{\Delta^\ddagger F - \Delta^\ddagger F^*}{k_B (T_\mathrm{vib} - T_\mathrm{bg})} .
\end{equation}
One can think of $\alpha$ as being a microscopic measure of efficacy, because it only applies to the role of the energy in a single vibrational mode.
$\alpha_d$, in contrast, more closely resembles the ensemble picture inherent to our model, because it assumes that all excess vibrational energy is equipartitioned over all degenerate modes which entails a loss of efficacy.
Figure~\ref{fig:panel-ch4}d then shows that, by the $\alpha_d$ metric, the efficacy of the symmetric stretch is higher than that of asymmetric stretches, which is in turn higher than that of the bending modes.
These results are in qualitative agreement with beam studies~\cite{Juurlink2005}.

Although excitation of bending modes is formally a less efficient way to stimulate methane dissociation (by the $\alpha_d$ metric), it is perhaps a more important reaction channel.
As noted before, at low $T_\mathrm{vib}$ and all else being equal the rate enhancement resulting from excited bending modes greatly exceeds that of the stretch modes.
This is in line with a suggestion already made by Juurlink et al.~\cite{Juurlink2005}
Moreover, in contrast to their beam experiments, also the efficacy of the bending modes is higher than that of the asymmetric stretch (which is however also degenerate, and lumped by construction in our model).
It is only at high $T_\mathrm{vib}$ that the efficacy of vibrational reaction stimulation of the bending mode decreases, asymptotically arriving at ${\sim} 0.2$.
At moderate $T_\mathrm{vib}$, however, we find $\overline{\alpha}_\mathrm{b} \approx 0.78$ which is close to the efficacy of 0.72 reported in beam studies~\cite{Juurlink2005}.

We have calculated $\overline{\alpha}$ values of the three considered modes for the three modeled Ni facets, collected in table~\ref{tab:ch4}.
These values were calculated from $\alpha_d$ between $T_\mathrm{vib} = 1000$~K and 2000~K for each mode and facet, respectively.
Due to the rater high noise in our $\alpha$ data, we have no consistently extrapolated $\alpha_\infty$ available.
Still, the values of $\overline{\alpha}$ provide some insight in the differences between the facets.
We can see that methane reactivity is similar on the (111) and (001) facets, and so is the vibrational efficacy.
In contrast to beam studies~\cite{Juurlink2005}, vibrational efficacy appears somewhat higher on (001), which may be tied to the relative overestimation of the dissociation barrier on this facet (or the underestimation on (111)).
When going to the (211) facet, efficacies drop.
Like for H$_2$ dissociation, this is in qualitative agreement with the FM model.
The FM model is however not quantitatively applicable to gas--surface reactions and cannot distinguish between different classes of vibrational excitation.

Conventionally, step sites dominate the reactivity of a catalyst in dissociation reactions, even when present in low concentrations~\cite{Dahl2000,Guo2018}.
In addition, small catalyst particles will usually expose a relatively large stepped surface area.
On one hand, this means that vibrational efficacy is always lower on  ``real life'' catalysts than on idealized flat surfaces.
On the other hand, it could be argued that a strong vibrational nonequilibrium affects the reactivity of flat terrace sites more strongly than steps.
As a result, the catalyst's surface as whole becomes more reactive, and the distinction between steps and terraces is diminished.
Along similar lines it has been argued that weakly binding catalysts benefit more from vibrational nonequilibrium than strong binding catalysts~\cite{Mehta2018,Engelmann2020}.

Our results can be directly assessed within the context of plasma catalysis.
Recently, vibrational nonequilibrium has been observed in methane plasma~\cite{Butterworth2020}.
In that work, overpopulated vibrationally excited states were found to exist shortly after a microwave pulse, and were well-described by a Boltzmann distribution.
These measurements thus justify some of the assumptions in our model.
In addition, the different vibrational modes were in equilibrium with one another.
As a result, excited bending modes were found in excess and, as noted earlier, would likely dominate chemistry with a lower overall efficacy.
However, Butterworth et al. only observed maximal vibrational nonequilibrium when $T_\mathrm{vib} < 700$~K, with $T_\mathrm{bg} \approx 300$~K.
This nonequilibrium was only maintained for a few $\mu$s before all degrees of freedom equilibrated at a high $T_\mathrm{bg}$~\cite{Butterworth2020}.
The residence time in typical plasma conversion experiments is in the order of seconds~\cite{Kim2016,Rouwenhorst2019,Uytdenhouwen2021}, which would mean that the gas is in vibrational equilibrium most of this time.
The very low $T_\mathrm{vib}$, relative to $T_\mathrm{bg}$, would therefore result in a limited decrease of the barrier that is only in play for a relatively short time.

On the flipside, recent kinetic analysis comparing catalytic dry reforming of methane in thermal and plasma setups indicate that certain plasma conditions reduce the apparent activation energy of the process~\cite{Sheng2020}.
The rate was found to be dominated by CH$_4$ dissociation, and plasma ignition decreased the apparent barrier by about 11~kcal/mol.
If we assume that the vibrationally-induced barrier reductions of figure~\ref{fig:panel-ch4}b are additive (and all modes on average contain equal amounts of energy), such an effect is already possible at a fairly modest nonequilibrium of $T_ \mathrm{vib} - T_\mathrm{bg} \approx 500$~K on a Ni(111) surface (but over 1000~K is needed on a stepped surface).
Note, however, that thermal catalysis was found to exhibit two regimes, depending on the temperature.
Most plasma-catalytic experiments were carried out at higher gas temperatures (${>}$700~K).
In such a regime, the thermal activation energy was already found to be reduced by about 9~kcal/mol compared to the low-temperature thermal reaction.
As a result, the effective plasma-induced barrier reduction at high gas temperature was only about 2~kcal/mol.
Consequently, rate enhancement was less than an order of magnitude.
If we take our predicted barrier reductions on the stepped Ni(211) surface as a reference, this would correspond to an excess vibrational temperature ${<}$250~K.
To wit, if we assume that the observed rate enhancement is due to vibrational nonequilibrium, it is still only a fairly minor effect.
Note, also, that microkinetic models predict that at sufficiently high $T_\mathrm{vib}$ other reaction steps become rate-determining~\cite{Engelmann2020}.

\section{Conclusions}

We have combined molecular dynamics simulations, free energy methods, and machine learning techniques to study the impact of a vibrational nonequilibrium on the catalytic dissociation of H$_2$ and CH$_4$.
We predict that vibrational energy very efficiently lowers the apparent free energy barrier of dissociative chemisorption at moderate vibrational nonequilibrium.
Vibrational efficacies of stretch modes tend to exceed those predicted by simpler theories: At low $T_\mathrm{vib}$ the vibrational efficacy of H$_2$ and the symmetric mode in CH$_4$ reaches almost 4, or almost an order of magnitude higher than predicted by the Fridman--Macheret model.
The vibrational efficacy of CH$_4$ bend modes is lower than that of stretch modes, but their total contribution to rate enhancement is larger due to their higher degeneracy.
The  vibrational efficacy of methane dissociation is lower on stepped surfaces, so that the impact of vibrational nonequilibrium may be dimished in real life catalytic setups.
Enhancement of H$_2$ dissociation, however, appears to be independent of the surface structure.

Comparison of our results with available experimental data suggests that the potential impact of vibrationally excited CH$_4$ on the overall catalytic process may be minor.
Most notably, limited available measurements point to a fairly weak and short-lived vibrational nonequilibrium in CH$_4$ plasma.
However, conversion rates have been reported to be higher in plasma-catalytic setups which have been attributed to vibrational nonequilibrium.
It must be kept in mind, however, that it remains quite difficult to properly disentangle the role of vibrationally excited molecules from the many other plasma-induced phenomena: Radicals, electric fields, energetic electrons, local hot spots, and so forth.
As we have shown again in this study, atomistic modeling allows for a highly controlled environment to asses such effects individually.

The conclusions presented here are contingent upon the quality of our simulations and our understanding of the vibrational nonequilibrium in methane plasma.
First, we have assumed that the ReaxFF force field is of sufficient quality, and that our ensemble-based model is adequate.
Second, additional experimental and modeling work will be necessary to obtain a more comprehensive insight into the vibrational populations, and to characterize surface kinetics inside a plasma-catalytic system.
As a result, we anticipate that further methodological advances, both theoretical and experimental, will better clarify the nature of vibrationally stimulated catalytic processes.

\ack
K.M.B. was funded as a junior postdoctoral fellow of the FWO (Research Foundation -- Flanders), Grant 12ZI420N.
The computational resources and services used in this work were provided by the HPC core facility CalcUA of the Universiteit Antwerpen, and VSC (Flemish Supercomputer Center), funded by the FWO and the Flemish Government.
HLDA calculations were performed with a script provided by G. Piccini.

\bibliography{bibliography}

\end{document}